# SOME METHODS OF ESTIMATING UNCERTAINTY IN ACCIDENT RECONSTRUCTION

Milan Batista

University of Ljubljana, Faculty of Maritime Studies and Transport, Slovenia, EU

March, 2010

**Abstract**
In this paper four methods for estimating uncertainty in accident reconstruction are discussed: total differential method, extreme values method, Gauss statistical method, and Monte Carlo simulation method. The methods are described and the program solutions are given

## 1 Introduction

A typical accident reconstruction problem is described by more or less sophisticated mathematical model i.e. a set of parameters connected by a set of equations describing a specific traffic accident situation ([5],[9]). A simplest example is speed from skids model where deceleration *a*, skid length *s* and velocity *v* are connected by $v = \sqrt{2as}$. Among the parameters there are input data, whose values is sought to be known and by which, when the equations are solved, the values of output data i.e. the answers of an expertise, are depend. The input data are ether measured (road grade, length of skid marks, positions of vehicles, etc), estimated from various published sources (friction coefficient, reaction time, etc) and/or also estimated from a testimony of witness. In other words the typical data in accident reconstruction practice are fuzzy and the question arise what is the effect of inaccuracy in input data on the determination of output data. This is not just a theoretical question. As it was pointed by Brach ([3]) the USA Federal Rules of Evidence include the demand that the known or potential rate of error of the technique should be determinate[1] and also that the critical cases should not be decided on results with high uncertainty.

Now, in the field of accident reconstruction the term *uncertainty* is synonym for variation, error, fluctuation, discrepancy etc ([3],[4],[15]). To make the term more precise let take that the model of an accident is described by set of equations

$$y_k = f_k(x_1, x_2, ..., x_n) \qquad (k = 1,..,m) \qquad (1)$$

or in vector form

$$\boldsymbol{y} = \boldsymbol{f}(\boldsymbol{x}) \qquad (2)$$

---

[1] This is not the case in Slovenia

where $\boldsymbol{x} = (x_1,...,x_n)$ is vector of input parameters and $\boldsymbol{y} = (y_1,...,y_m)$ is vector of output parameters. The model (1) can be easily implemented as program subroutine. The example of such a subroutine which calculates the critical speed ([5]) is given in Figure 1.

```fortran
      subroutine model02( x, y, stat)
!
!         Critical speed - SAE 950354
!
      implicit none
      real, intent(in)  :: x(*)  ! Input parameters
      real, intent(out) :: y(*)  ! Output parameters
      integer, intent(out) :: stat  ! Error indicator 0=OK
      real, parameter :: g = 9.81
      real :: mu, r, p, v
      stat = 0
!
!         Unpack
      mu = x(1)         ! average decleration factor
      r  = x(2)         ! radius
      p  = x(3)/100.0   ! grade
!
!         Calculate
      v  = r*g*(mu + p)/(1.0 - mu*p)
      if (v <= 0.0) then
          stat = 1
          return
      endif
      v  = sqrt(v)
!
!         Pack
      y(1) = v*3.6
!
      end subroutine model02
```

**Figure 1. A Fortran90 program implementing the critical speed model.**

Let each of the input parameters be described in the form ([12])

$$x_j = x_{j0} \pm \Delta x_j \qquad (j = 1,...,n) \qquad (3)$$

where $x_{j0}$ some best estimate for $x_j$ and $\Delta x_j \geq 0$ is its error or uncertainty. Thus it is somehow expected the true value for $x_j$ lies in the interval $x_{j0} - \Delta x_j \leq x_j \leq x_{j0} + \Delta x_j$. Given the relation (1) and input parameters in the form (3) one now asks what effect has inaccuracy in $\boldsymbol{x}$ on the determination of $\boldsymbol{y}$ i.e. one ask for the output parameters in the form

$$y_i = y_{i0} \pm \Delta y_i \qquad (i = 1,...,m) \qquad (4)$$

The determination of a best estimate $y_{i0}$ for $y_i$ and its uncertainty $\Delta y_i \geq 0$ for the given model (1) is the task of uncertainty analysis. There are several methods found in the literature ([3],[4],[14],[15]) available for uncertainty analysis which can be divided into two classes:

- non-statistical methods (total differential method, extreme values method)

- statistical methods (Gauss method, Monte Carlo method)

These methods will be shortly discussed in the following sections.

## 2  Non-statistical methods

In non-statistical method of uncertainty analysis one start with the assumption that the value of each input parameter is given in a range $x_j \in \left[ x_{j\min}, x_{j\max} \right]$. This may be put in the from (3) if one at hoc estimate the best value for $x_j$ as arithmetic mean and uncertainty as half of range

$$x_{j0} = \frac{x_{j\min} + x_{j\max}}{2} \qquad \Delta x_j = \frac{x_{j\max} - x_{j\min}}{2} \qquad (5)$$

The question is now how to determine a range of values for $y_i \in \left[ y_{i\min}, y_{i\max} \right]$ for a given ranges of input parameters and the model (1). Here it must be stressed that if one assumes that the ranges for input parameters are absolute so the answer on ranges of output parameters is absolute too. Thus by this methods the answer for uncertainty is: if ranges for input parameters are $x_j \in \left[ x_{j\min}, x_{j\max} \right]$ then the ranges for output parameters are $y_i \in \left[ y_{i\min}, y_{i\max} \right]$. Two methods will now be shortly discussed: the total differential method and the extreme values method.

### 2.1  Total Differential Method

The discussion and examples of this method are given in [7]. The basic idea of the method is the following ([6]). If instead of 'best' value of $x_0$ one use inaccurate value $x_0 + \delta x$ then the correspondent value of $y$ differs from best value $y_0 = f(x_0)$ by $\delta y = f(x_0 + \delta x) - f(x_0)$. By Taylor theorem one than has

$$\delta y_i = f_i(x_{10},...,x_{n0}) + \frac{\partial f_i}{\partial x_1} \delta x_1 + ... + \frac{\partial f_i}{\partial x_n} \delta x_m + ... - f_i(x_{10},...,x_{n0})$$

where the partial derivatives are calculated at nominal values. Now if the uncertainties $\Delta x_j$ are small and $|\delta x_j| \leq \Delta x_j$ then the uncertainties in $y_i$ may be approximated as

$$\delta y_i \approx \frac{\partial f_i}{\partial x_1} \delta x_1 + ... + \frac{\partial f_i}{\partial x_n} \delta x_n$$

Since absolute value of uncertainties is needed, one must estimate the above

$$|\delta y_k| \approx \left| \frac{\partial f_k}{\partial x_1} \delta x_1 + ... + \frac{\partial f_k}{\partial x_n} \delta x_n \right| \leq \left| \frac{\partial f_k}{\partial x_1} \right| \Delta x_1 + ... + \left| \frac{\partial f_k}{\partial x_n} \right| \Delta x_n$$

From this one my set the estimate for uncertainty on output variables $y_i$ as

$$\Delta y_i \leq \left|\frac{\partial f_i}{\partial x_1}\right|\Delta x_1 + ... + \left|\frac{\partial f_i}{\partial x_n}\right|\Delta x_m \quad (6)$$

```fortran
subroutine TDM( model, nx, x0, dx, ny, y0, dy)
    integer, intent(in) :: nx, ny
    real, intent( in) :: x0(nx), dx(nx)
    real, intent(out) :: y0(ny), dy(ny)
    external model
    intrinsic abs, max, matmul
    integer :: k, n, nn, stat
    real :: xx0, dxx
    real :: xx(nx), ymin(ny), ymax(ny), dfdx(ny,nx)
    call model( x0, y0, stat)
     xx(1:nx) = x0(1:nx)
    do k = 1, nx
        xx0 = x0(k)
        dxx = dx(k)
        if (abs(dxx) <= 0.0) dxx = max(0.025*xx0,0.01)
        xx(k) = xx0 - dxx
        call model ( xx, ymin, stat)
        xx(k) = xx0 + dxx
        call model( xx, ymax, stat)
        xx(k) = xx0
        dfdx(1:ny,k) = abs((ymax(1:ny) - ymin(1:ny))/(2.0*dxx))
    end do
    dy(1:ny) = matmul(dfdx(1:ny,1:nx),dx(1:nx))
end subroutine
```

**Figure 2. A Fortran90 subroutine which implement the total differential method using central finite difference approximation of partial derivatives**

The main weakness of the method is that one must calculate the partial derivatives of (1). For simple models this can be done analytically, but for complex model this must be done numerically by using finite differences. For example, the central finite difference approximation of derivative is ([4])

$$\left(\frac{\partial f_i}{\partial x_k}\right)_0 \approx \frac{f_i(x_{10},...x_{k0}+\delta x_k,...x_{n0}) - f_i(x_{10},...x_{k0}-\delta x_k,...x_{n0})}{2\delta x_k} \quad (7)$$

which is of order $O(\delta x_k^2)$. For practical calculation one can for $\delta x_k$ use provided data i.e. one can set $\delta x_k = \Delta x_k$. However when $\Delta x_k = 0$ one can select any value for calculation of a partial derivative because in this case then there is no uncertainty in $x_k$ and this input parameters drops from the formula (6). Note that the partial derivatives can be used to establish the parameter dependence. If $\left(\frac{\partial f_i}{\partial x_k}\right)_0 = 0$ then $y_i$ is independent of $x_k$.

The Fortran code which implements the described methods is given in Figure 2.

## 2.2 Extreme Values Method

This method is in essence the minimization/maximization problem with simple bounds. Let the definition region of $f(x)$ be

$$\Omega = [x_{1\min}, x_{1\max}] \times ... \times [x_{n\min}, x_{n\max}]$$

then one has to find the points $x \in \Omega$ where $f(x)$ has the global extremes i.e. global minimum $\min_{x \in \Omega} f(x)$ and the global maximum $\max_{x \in \Omega} f(x)$. If the function $f(x)$ is continuous then the local extremes are found by solving the set of equations obtained by the first partial derivatives inside region $\Omega$

$$\frac{\partial f_i}{\partial x_j} = 0 \quad (i = 1,...,m;\ j = 1,...,n)$$

However to find global extremes one must find the local extremes also on the boundary of the region $\Omega$ to ([11]).

```fortran
subroutine EVM( model, nx, xmin, xmax, ny, ymin, ymax)
    integer, intent(in) :: nx, ny
    real, intent( in) :: xmin(nx), xmax(nx)
    real, intent(out) :: ymin(ny), ymax(ny)
    external model
    intrinsic btest, max, min
    integer :: k, n, nn, stat
    real :: x(nx), y(ny)
    do n = 0, 2**nx - 1
        do i = 1, nx
            if (btest(n,i-1)) then
                x(i) = xmax(i)
            else
                x(i) = xmin(i)
            endif
        enddo
        call model ( x, y, stat)
        if (stat /= 0) cycle
        if (n == 0) then
            ymin(1:ny) = y(1:ny)
            ymax(1:ny) = y(1:ny)
        else
            do i = 1, ny
                ymin(i) = min(y(i),ymin(i))
                ymax(i) = max(y(i),ymax(i))
            enddo
        endif
    enddo
end subroutine
```

**Figure 3**. **A Fortran90 program implementing the min/max variant of extreme value method**

In special case when $f(x)$ is linear then the extreme values are to be found in a corner points of the region $\Omega$. This can be done simply by taking in calculation all possible variations of $x$ min/max values. For $n$ input variables there are $2^n$ possible variations. For example when $n = 16$ one obtains 65536 variations. Note that the particular variation of min/max values for $x$ can be constructed on the base of the binary representation of the value of $2^j$, $(0 \leq j \leq n-1)$ where 0 is selected for min and 1 for max value of corresponding variable $x_{j+1}$. When $f(x)$ is weakly-nonlinear or it is linearized and also the uncertainties $\Delta x_j$ are small this method can also be used to estimate extremes of $f(x)$. The Fortran code which implements the described methods is given in Figure 3.

Before proceed it must be emphases that the min/max variant of the method was criticized by several authors. Thus Kost and Werner ([8]) criticized the variant of the method when one use in the first calculation all min and than all max values by stating that the chance that all of parameters being either max or min should be virtually zero and also that the method is to cumbersome and time-consuming. The similar objection is also pointed out by Brach ([3]) stating that the likehood of indepedned variables to simultaneously reach the extreme values is not taken into account so the results can be unrealistic.

## 3 Statistical methods

In statistical methods of uncertainty analysis the model parameters are taken as random variables. Thus the best estimate for $x_i$ is taken to be the mean value $\mu_{x_i}$ and the measure for uncertainty $\Delta x_i$ is a multiple of standard deviation $\sigma_{x_i}$. Two methods will be shortly discussed: the Gauss method and Monte Carlo method.

### 3.1 Gauss Method

As it can be shown if $x$ is random vector with probability distribution $\varphi(x)$ then the mean value or expected values of output parameters are given by ([10])

$$\mu_{y_i} = E(y_i) = \iint \ldots \int f_i(x_1,\ldots,x_n)\varphi(x_1,\ldots,x_n)dx_1\ldots dx_n \qquad (8)$$

and the variance is defined as

$$\sigma_{y_i}^2 = \mathrm{Var}(y_i) = E\left\{\left[y_i - E(y_i)\right]^2\right\} \qquad (9)$$

The general evaluation of the mean and variance by the above formulas is difficult as it depends on complexity of the functions $f_k(x)$ and probability distribution $\varphi(x)$. However if the functions $f_k(x)$ are linarized then the problem become easy. By expanding the functions (1) into the Taylor series around the mean values of input parameters one obtains

$$y_i = f_i(x_1,...,x_n) \approx f_i(\mu_{x_1},...,\mu_{x_n}) + \frac{\partial f_i}{\partial x_1}(x_1 - \mu_{x_1}) + ... + \frac{\partial f_i}{\partial x_n}(x_n - \mu_{x_n}) + ... \quad (10)$$

where partial derivatives are evaluated at mean values. Substituting this into (8) and take taking into account that $\iint...\int \varphi(x_1,...,x_n)dx_1...dx_n = 1$ and $\mu_{x_k} = \iint...\int x_k\varphi(x_1,...,x_n)dx_1...dx_n$ one obtains

$$\mu_{y_i} \approx f_i(\mu_{x_1}, \mu_{x_2},...,\mu_{x_n}) \quad (11)$$

Similarly by substituting (10) into (9) and take (11) into account one arrives to

$$\sigma_{y_i}^2 \approx \sum_{k=1}^{n}\left(\frac{\partial f_i}{\partial x_k}\right)^2 \text{Var}(x_k) + 2\sum_{\substack{j=1\\j\neq k}}^{n}\sum_{k=1}^{n}\left(\frac{\partial f_i}{\partial x_j}\right)\left(\frac{\partial f_i}{\partial x_k}\right)\text{Cov}(x_j,x_k) \quad (12)$$

where $\text{Cov}(x_j,x_k) = E\left[(x_j - \mu_{x_j})(x_k - \mu_{x_k})\right]$ If future is assume that the input parameters are independent i.e. that $\text{Cov}(x_j,x_k) = 0$ one obtain so called propagation of error formula

$$\sigma_{y_k}^2 \approx \left(\frac{\partial f_k}{\partial x_1}\right)^2 \sigma_{x_1}^2 + \left(\frac{\partial f_k}{\partial x_2}\right)^2 \sigma_{x_2}^2 + ... + \left(\frac{\partial f_k}{\partial x_n}\right)^2 \sigma_{x_n}^2 \quad (13)$$

The program realization of the described method is almost the same as the total differential method (Figure 2) except that calculation of of vector `dfdx(1:ny,k)` is replaced by

```
dfdx(1:ny,k) = ((ymax(1:ny) - ymin(1:ny))/(2.0*dxx))**2
```

Future discussions and examples of using this method can be found in [2] and [13].

### 3.2 Monte Carlo Method

The basic idea behind the Monte Carlo method is to supply deterministic model with random generated data. By repeated computation one obtains the sample which can then be used to estimate the statistics of the output parameters i.e. sample mean value

$$\overline{y}_i = \frac{y_{i_1} + y_{i_2} + ... + y_{i_N}}{N} \quad (14)$$

and sample variance

$$s_i = \frac{1}{N-1}\sum_{k=1}^{N}(y_{i_k} - \overline{y}_i)^2 \quad (15)$$

where *N* is the number of calculations. Examples of using the method can be found in [1][8][14][16]. For practical use of method the question of which probability distribution one should take to the particular input parameter must be answered. As was observed by Woods in accident reconstruction the distribution of the majority of input parameters are unknown so he suggests most conservative distribution should be used: uniform probability distribution ([16]).

```fortran
      subroutine MCM( model, nr, nx, xmin, xmax, ny, ybar, ysdv)
         integer, intent(in) :: nr ! Number of runs
         integer, intent(in) :: nx, ny
         real, intent( in) :: xmin(nx), xmax(nx)
         real, intent(out) :: ybar(ny), ysdv(ny)
         external model
         intrinsic sqrt, random_number, random_seed
         integer :: k, n, nn, stat
         real :: t, x(nx), y(ny)
         call random_seed()
          ybar(1:ny) = 0.0; ysdv(1:ny) = 0.0
          nn = 0
         do n = 1, nr
            do k = 1, nx
               call random_number(t)
               x(k) = xmin(k) + (xmax(k) - xmin(k))*t
            enddo
            call model( x, y, stat)
            if (stat /=0) cycle
            nn = nn + 1
            ybar(1:ny) = ybar(1:ny) + y(1:ny)
            ysdv(1:ny) = ysdv(1:ny) + y(1:ny)**2
         enddo
         if (nn <= 1) return
         ybar(1:ny) = ybar(1:ny)/nn
         ysdv(1:ny) = sqrt((ysdv(1:ny) - nn*ybar(1:ny)**2)/(nn-1))
      end subroutine
```

**Figure 4. A Fortran90 subroutine which implement the Monte Carlo simulation together with basics statistics calculation**

The program segments which performs Monte Carlo simulation with assumptions of uniform probability distribution of input parameters and also calculate the basic statistics is given on Figure 4.

## 4   Conclusion

From the present treatment, one can conclude that there is no universal method for uncertainty analysis. All of the present methods has some disadvantage or are based on some assumptions. Typically, the total differential method is based on assumption that the uncertainties of input parameters are small. The same is assumed in Gauss method with addition that one must in advance assume a interpretation of input parameters uncertainty and probability distribution. This last is also needed for Monte Carlo method.

It must be stressed that the uncertainty analysis does not give the answers to the questions such as how one should treat single value, how should one treat the published range of data, how should one interpret min and max value,  are all input parameters independent .etc.. Consequently the interpretation of the results of uncertainty analysis must always be in the form of implication:. if the assumption about input varables are  satisfied then the results are within such and such limits.The described method are thus a valuable tools to answer some questions about uncertainty but one should bear in mind that each accident is unique and should be treated as such.